\input FEYNMAN
\documentstyle[11pt]{article}
\title{~~~~ \\ QED RADIATIVE CORRECTIONS \\ TO
THE DECAY $\pi^{0} \rightarrow e^{+}e^{-}$}
\author{George Triantaphyllou\\
Department of Physics, Yale University, New Haven, Ct. 06520}
\begin{document}
\setlength{\baselineskip}{24pt}
\maketitle
\setcounter{page}{0}

{}~ \\

\begin{abstract}
 In view of the recent interest in the decays of mesons into a pair of light
leptons, a computation of the QED
radiative corrections to the decay of $\pi^{0}$ into an
electron-positron pair is presented here.
The results indicate that the peak value of the
differential decay rate is reduced by about $50 \%$,
because of soft-photon radiation. The number of $e^{+}e^{-}$ pairs having
invariant masses in an energy bin of 0.5 MeV centered around $m_{\pi}$ is
found to be about
$20 \%$ smaller than the one it would be if radiative corrections
 were neglected.
\end{abstract}

\pagebreak

The decay of $\pi^{0}$ into a pair of electrons via two virtual photons
(see Fig.\ref{fig:diagram})
was studied long ago \cite{nuevo}. The authors of Ref.\cite{nuevo} calculated
the absorptive part of that decay amplitude, in order to derive a unitarity
bound for the branching ratio
\begin{equation}
 B \equiv \frac{\Gamma(\pi^{0} \rightarrow e^{+}e^{-})}
{\Gamma(\pi^{0} \rightarrow \gamma \gamma)},
\end{equation}
\noindent using as input the experimentally measured value for
$\Gamma(\pi^{0} \rightarrow \gamma \gamma)$.
They found that $B \geq 4.7 \times 10^{-8}$, an exact result. Similar
exact unitarity
bounds can be derived for the decays of heavier mesons into lepton-antilepton
($ l^{+}l^{-}$ ) pairs.

To complete the above calculation though, one would also
 need the dispersive part
of the amplitude. The knowledge of the exact branching ratio
 is essential, because
it could provide us with a test of new physics \cite{Berg}. To be more precise,
the decay of a meson to a $l^{+}l^{-}$ pair could be influenced
by the existence of point-like
effective 4-Fermi interactions between the constituent
quarks of the meson and the final two leptons. Such interactions exist in
technicolor models, in models of composite quarks and leptons, {\em etc}.
In technicolor models, for example, extended-technicolor forces couple
not only technifermions to ordinary fermions, but also ordinary fermions
to themselves.

The dispersive parts of the two photon contribution to the mesonic
decay amplitudes are difficult to calculate, however, since
in that case
 the two intermediate photons are off-shell; there exists presently
no exact model describing the non-perturbative meson dynamics needed to make
this calculation. In addition, the problem is even more complex in the case
of the pion. This is because this meson plays
also the role of a Goldstone-boson, arising
from the spontaneous breakdown of
the $ SU(2)_{L} \times SU(2)_{R} $ chiral symmetry of QCD, and the
assumption that it is
just a bound state of two quarks must be viewed with caution.

\begin{figure}[t]
\vspace{6cm}
\begin{picture}(6000,6000)
%
%
%
%
%
\end{picture}
\caption{The decay $\pi^{0} \rightarrow e^{+}e^{-}$. The 4-momenta
 of the various particles are given inside parentheses.}
\label{fig:diagram}
\end{figure}

There have been several attempts to estimate the dispersive part of the decay
amplitude, using a vector dominance model \cite{Lits}, leading to
$B = 1 \times 10^{-7}$,
a nucleon loop model
 \cite{Pratap}, leading to $B = 1.4 \times 10^{-7}$,
 an  extreme-non-relativistic model \cite{Berg,Berg1},
leading to $B = 6.2 \times 10^{-8}$,
a relativistic bag model \cite{Margolis}, leading to
 $B \approx 1 \; \times 10^{-7}$, and a chiral Lagrangian approach
\cite{Savage}, leading to  $B = 7 \pm 1 \; \times 10^{-8}$.
  Early experimental
results  \cite{Mis} gave $B = 1.8 \pm 0.6 \times 10^{-7}$, which was somewhat
large, and caused some controversy over the theoretical predictions
\cite{Berg2}.
Nevertheless, the latest experiments in Fermilab
 \cite{Fermi} and Brookhaven \cite{Brook}
, which give $B = 6.9 \pm 2.8 \times 10^{-8}$ and
$B = 6.0 \pm 1.8 \times 10^{-8}$ respectively, indicate
 that the dispersive part of
the amplitude is indeed smaller that the absorptive part.

Another issue relevant to the decay $\pi^{0} \rightarrow e^{+}e^{-}$ is
the one related to final-state QED radiative corrections. The fact that
$m_{\pi} \gg m_{e}$
 makes the two electrons in the final state quite energetic, and radiative
effects can therefore be quite important. Similar instances, where QED
radiative corrections are large, have already been observed
in different mesonic decays into leptons \cite{other}.
Radiation of soft photons can
alter substantially the decay profile of the pion, reducing considerably
the peak differential decay rate of this process, and obviously affecting
the branching ratio B. This paper is focused on these
corrections, and their treatment is done independently of the method used
to calculate the non-radiatively-corrected decay rate.

An early calculation
of QED radiative corrections to the decay of a meson (a kaon) into two leptons
via two virtual photons, to first order in the fine structure constant
$\alpha$, appeared in Ref.\cite{Patil}. Later, a similar calculation appeared
 \cite{Bergrad} for $\pi^{0} \rightarrow e^{+}e^{-}$, the decay studied here.
That calculation was also limited to  first order in $\alpha$.
In such a process, however, $e^{+}e^{-}$-invariant-mass cuts must be very
strict, as much as experimental resolution allows, in order
 to minimise the background effects, an issue that will
be discussed later in the paper. And it was made clear in Ref.\cite{Bergrad}
that the higher orders in the perturbative expansion in $\alpha$
influence considerably the differential decay rate,
if the  $e^{+}e^{-}$-invariant-mass resolution
   becomes  smaller than about $1 \%$ of the pion mass. This is so
 because, in that case,
 the first-order result already gives a larger-than-$20 \%$ correction.
The present experimental
resolution for the $e^{+}e^{-}$ invariant mass being on the order of
$0.5 \; {\rm MeV} \approx 0.4 \% \;m_{\pi}$ \cite{Zeller}, an improved
calculation of the
radiative corrections for this decay is needed, in order to include higher
orders in $\alpha$.

The following analysis is similar to the one usually applied to calculate
the initial-state radiative corrections to meson production via
electron-positron collisions \cite{meson}.
The differential decay rates, denoted by $P(x)$ below,
where $x$ is the invariant mass of either the pion or the $e^{+}e^{-}$ pair,
 are defined here to be equal to
$\frac{d\Gamma(\pi^{0} \rightarrow e^{+}e^{-})}{dx}$.
Assuming a relativistic Breit-Wigner shape for the pion resonance, the
result for $P(s)$ to zeroth order in the fine structure
constant $\alpha$ is
\begin{equation}
P_{0}(s) =
 P_{0}(m^{2}_{\pi})\;\frac{ s \Gamma^{2}_{\pi}}{(s - m^{2}_{\pi})^{2} +
\Gamma^{2}_{\pi}m^{2}_{\pi}} ,
\label{eq:Lorenz}
\end{equation}

\noindent where $\sqrt{s}$ is the invariant mass of $\pi^{0}$,
 $m_{\pi} = 134.9$ MeV is its mass, and
$\Gamma_{\pi} \approx 8$ eV its natural width
\footnote{This is a different approach from the one in
Ref.\cite{Bergrad}, where
$\frac{d\Gamma(\pi^{0} \rightarrow e^{+}e^{-})}{ds}$,
 the quantity corresponding to $P_{0}(s)$, is
taken to be proportional to $\delta(s - m^{2}_{\pi})$.}.
The pion natural width is
so small, that any possible dependence of $P_{0}$ on s coming from the
form factor associated to the decay $\pi^{0} \rightarrow e^{+}e^{-}$
is neglected.

Generally, if the experimentally measured quantity is the invariant mass
$\sqrt{s^{'}}$ of the $e^{+}e^{-}$ pair,
the non-corrected and corrected differential decay rates,
 denoted by $P_{0}(s^{'})$ and $P_{\rm rad}(s^{'})$ respectively,
 can be both expressed as follows:

\begin{equation}
P_{0, {\rm rad}}(s^{'})  =
 \int_{s^{'}}^{2s^{'}} ds P_{0}(s) f_{0, {\rm rad}}(s-s^{'})
\label{eq:integr}
\end{equation}

\noindent
The distribution $f_{0, {\rm rad}}(s-s^{'})$ is
subject to the requirement that
\begin{equation}
\; \int_{s^{'}}^{2s^{'}} ds f_{0, {\rm rad}}(s-s^{'}) = 1.
\end{equation}
 In a reference frame
where the two final electrons have equal and opposite 3-momenta,
  the relation $s-s^{'} = 4m_{e} \omega$ holds, where $\omega$
is the total energy of the radiated photons, and $m_{e}$ is the mass
of the electron. The upper bound
 $ (2 s^{'}) $ of the above
integration is placed  in order to avoid the pair creation
 of $e^{+}e^{-}$ pairs
having invariant mass $s^{'}$ and originating from the radiated photons,
since  they could be misidentified as being  $e^{+}e^{-}$ pairs
 having  the same invariant mass and originating  from the pion.
 More on this upper bound will be discussed shortly.

 The differential decay rate $P_{0}(s^{'})$ is associated to
 the distribution
\begin{equation}
 f_{0}(s-s^{'}) = \delta(s-s^{'})
\end{equation}
 \noindent
because, to zeroth order in $\alpha$, $s = s^{'}$.
The effect of the radiation of soft photons is the "smearing" of this
$\delta$-function, even though the resulting
differential decay rate still retains an
integrable singularity at $ s = s^{'} $. The convolution of the
 Breit-Wigner shape
of Eq.(\ref{eq:Lorenz}) with the adequate
"smearing" distribution $f_{{\rm rad}}(s-s^{'})$, the calculation of which is
the main purpose of the present paper,
 gives us the radiatively corrected  differential decay rate
$P_{{\rm rad}}(s^{'})$.

The pion is assumed to be produced hadronically, as given, for instance,
 by the hadronic reaction
$K^{+} \rightarrow \pi^{0} \pi^{+}$, and therefore
 only final-state QED radiative corrections
are considered. Furthermore, it is assumed that all radiated photons, no
 matter what their energy is, escape detection.
The calculation presented here is restricted to
the soft-photon resummation technique,
since this is known to give the bulk of the radiative corrections. The
final result, applicable
only to $e^{+}e^{-}$ invariant masses very close to the pion mass,
  is non-perturbative, since the leading
logarithms
of diagrams of all orders in $\alpha$ are summed,
in a process well known as "exponentiation" \cite{Yennie}.

In this context,  $P_{{\rm rad}}$ can be expressed as an infinite
sum of individual differential decay rates
 $P(\pi^{0} \rightarrow e^{+}e^{-} + n \gamma)$, each having $n$
real soft photons
radiated from the $\pi^{0}$:

\begin{equation}
 P_{ {\rm rad} }(s^{'}) = \sum_{n = 0}^{\infty}
P( \pi^{0}
\rightarrow  e^{+}e^{-} + n \gamma).
\end{equation}

\noindent Intuitively, this infinite sum has the meaning that in the
soft-photon (classical) limit, where
radiation loses its particle-like properties,
 two final states containing $n$ and $n + 1$ photons
respectively are undistinguishable, so a summation over all possible numbers
of photons is required. In such an analysis, each of the quantities
 $P(\pi^{0} \rightarrow e^{+}e^{-} + n \gamma)$ contain the virtual
photon corrections needed to cancel the infrared divergencies. Moreover,
the $n$ soft photons are
considered to be independant from each other. For this to be true,
their emission
must not influence considerably the motion of the electrons, so they
must have an energy much smaller than the electron. Therefore, the relation
$s - s^{'} \ll 4m^{2}_{e}$ must hold, and then the logarithms of the
form  $\ln{(s/s^{'} - 1)}$ dominate the QED perturbative expansion,
and are summed to all orders in $\alpha$.

The appearance of these logarithms
is closely connected to the cancellation of the infrared divergencies.
In order to see this, it is assumed that
the emission of  soft real photons, of energy less than $\omega_{0}$,
 is used in order to calcel the infrared divergencies
associated to the virtual photons.
Then the calculation involving soft real and virtual photons
 contains logarithms of the form
$\ln{\left(\frac{\omega_{0}}{\sqrt{s^{'}}}\right)}$. On
the other hand, the hard-real-photon part of the calculation, involving
photons of energy larger than $\omega_{0}$,
 contains logarithms of the form
$\ln{\left(\frac{s - s^{'}}{\sqrt{s^{'}}\omega_{0}}\right)}$.
The addition of these two contributions
cancels the dependance of the radiative corrections on $\omega_{0}$,
leaving logarithms of the form $\ln{(s/s^{'} - 1)}$ in the final result.

The soft photons are assumed to be radiated only by the two real final
electrons. This is because
the pion decay
cannot proceed via the emission of three photons, because of C-conservation,
 so no photon can come from there.
Moreover, the effect of
photons emitted by the virtual electron, denoted by $e^{*}$
  on the right side of the
 "$\gamma \gamma e^{*}$" triangle (see Fig.\ref{fig:diagram}),
 is neglected.
 Intuitively, this happens because the radiated soft-photons,
 which give the largest corrections, correspond to
 large spatial scales, and are therefore associated to the two
final real electrons. The virtual electron, however,
is far off-shell, since the typical momenta flowing
in the triangle of Fig.\ref{fig:diagram} are on the order of $m_{\pi}$,
 and it is associated to
small spatial scales, so its contribution to these corrections is negligible.

According to the previous discussion,
only radiative corrections pertinent to the two final electrons
are considered, and the result is, according to the analysis of
Ref. \cite{Yennie,Berends}

\begin{equation}
 f_{{\rm rad}}(s - s^{'})
=  \beta \frac{(s - s^{'})^{\beta - 1}}
{s^{' \beta}}\Delta^{1}_{{\rm rad}} + \frac{\Delta^{2}_{{\rm rad}}}{s^{'}}
\end{equation}

\noindent with

\begin{equation}
 \;\; \beta = \beta(s^{'})  =  \frac{2 \alpha}{\pi}
\left(\ln{(s^{'}/m^{2}_{e})} - 1 \right).
\label{eq:rad}
\end{equation}

\noindent
This expression gives a reasonable estimate
 only in the relativistic limit $s{'} \gg m^{2}_{e}$, a relation that
obviously holds for $s^{'}$ near the pion mass.
Moreover, this estimate is even more accurate
when $\beta \ll 1$. Here $ \beta(m^{2}_{\pi}) \approx 0.047$.

The quantities $\Delta^{1,2}_{{\rm rad}}$
correspond to
radiative corrections that can be treated perturbatively. Among various
other terms, they contain powers of $\beta$, i.e. basically terms of the form
$(\alpha \ln{(s^{'}/m^{2}_{e})})^{n}$, that were left out of the
exponentiation procedure.
 The effects of vacuum polarization of the two virtual
photons are also assumed to have been absorbed in $\Delta^{1,2}_{{\rm rad}}$.
To zeroth order in $\alpha$,
$\Delta^{1}_{{\rm rad}} = 1$ and $\Delta^{2}_{{\rm rad}} = 0$.
In the following, the approximation
\begin{equation}
\Delta^{1}_{{\rm rad}}  \approx  1 \nonumber
\end{equation}
 \noindent and
\begin{equation}
\Delta^{2}_{{\rm rad}}  \approx  0
\end{equation}
\noindent is used, i.e. only
the factor multiplying $\Delta^{1}_{{\rm rad}}$,
the non-perturbative part of the calculation,
is considered, because it is expected to give the bulk of the
 radiative corrections. In view of the magnitude of $\beta$, the error induced
by this approximation should be on the order of $5 \%$.

The
distribution $f_{{\rm rad}}(s-s^{'})$ contains the
exponentiated logarithms, which give the leading
QED correction to the decay profile for $s \approx s^{'}$.
Since $\beta < 1 $,
$f_{{\rm rad}}(s-s^{'})$ has an integrable
singularity at $s = s^{'}$.
It is worth noting that this form of $f_{{\rm rad}}(s-s^{'})$ implies the
assumption that the logarithms $\ln{(s/s^{'} - 1)}$ dominate the
QED perturbative expansion of the radiative corrections. This ceases to
be true when, in the integral of Eq.(\ref{eq:integr}),
  the integration variable $s$ approaches the upper bound ($2s^{'}$)
of the integration. Moreover, in that case, interference effects between
this process and the main background process, which will be discussed
shortly, become non-negligible.
However, the present form of the "smearing"
distribution $f_{{\rm rad}}(s-s^{'})$, and the smallness
of $\beta$ in Eq.(\ref{eq:rad}), makes
the integral of Eq.(\ref{eq:integr}) completely dominated by the region of
$s \approx s^{'}$. Therefore, effects coming from a large $s - s^{'}$
difference are
not expected to influence the final result considerably.

\begin{figure}[t]
\vspace{4.5in}
\caption{ The integrable singularity of the quantity
 $ \frac{2 s^{1/2}}{P_{0}(m^{2}_{\pi})} P_{0}(s)f_{{\rm rad }}(s-s^{'})$,
evaluated at $s^{'} = m^{2}_{\pi}$.}
\label{fig:differ}
\end{figure}

In Fig. \ref{fig:differ},
the quantity
$\frac{2 \sqrt{s}}{P_{0}(m^{2}_{\pi})}P_{0}(s)f_{{\rm rad}}(s-s^{'})$,
evaluated at $s^{'} = m^{2}_{\pi}$,
is plotted as a function of the pion invariant mass minus $m_{\pi}$, i.e.
 $\sqrt{s} - m_{\pi}$. The singularity at $s = s^{'}$  shows clearly.
The area under the curve that is partially shown in Fig.\ref{fig:differ},
contained in the region
$0 \; < \; \sqrt{s} - m_{\pi} \; < \; m_{\pi}$ , is equal to
$P_{{\rm rad}}(m^{2}_{\pi})/P_{0}(m^{2}_{\pi})$.

It is not difficult to see from Eq.(\ref{eq:integr}) and (\ref{eq:rad})
that, for $\Gamma_{\pi} \ll m_{\pi}$ and $\beta \ll 1$,

\begin{equation}
P_{{\rm rad}}(m^{2}_{\pi}) \approx P_{0}(m^{2}_{\pi})
\left(\frac{\Gamma_{\pi}}{m_{\pi}}\right)^{\beta},
\label{eq:central}
\end{equation}

\noindent with $\beta$ evaluated at $s^{'} = m^{2}_{\pi}$.
The ratio $\frac{\Gamma_{\pi}}{m_{\pi}}$ appears naturally, as being
the characteristic energy scale over which the pion differential
decay rate varies considerably, divided by
the largest energy scale of the process.
 The dependance of the right-hand side of Eq.(\ref{eq:central})
 on $m_{e}$ comes only
through the parameter $\beta$, which contains the term
$\ln{(s^{'}/m^{2}_{e})}$,
responsible for the collinear singularity in the limit
$m_{e} \rightarrow 0$. The fact that $m_{e}$ does not appear anywhere else
in the final result is attributed to the factorisation
of collinear singularities, a well-known property of QCD, which is also
shared by QED \cite{Berends}.
For the given values of $\beta(m^{2}_{\pi})$,
$\Gamma_{\pi}$ and $m_{\pi}$, Eq.(\ref{eq:central}) leads to the relation

\begin{equation}
P_{{\rm rad}}(m^{2}_{\pi}) \approx 0.46 \; P_{0}(m^{2}_{\pi}).
\end{equation}

Therefore, the peak value of the radiatively corrected
differential decay rate  is
about $50 \%$ smaller than the one without radiative corrections
\footnote{Similar examples can be found in
 connection with the production cross-section of heavy mesons
via $e^{+}e^{-}$ colliding beams, where the
QED radiative corrections are predicted to have a $50\%$ effect \cite{meson}.
 In those cases, however, in contrast to the analysis performed
in this paper, the characteristic energy scale over which the
production cross-section varies considerably is determined not by
the width of the produced resonance, but by the energy
spread of the colliding beams, which is much larger than the width.}.
The magnitude of this effect shows that,
 for $e^{+}e^{-}$ invariant masses
very close to  $m_{\pi}$, a
calculation of these corrections using perturbation theory up to
first order in $\alpha$ would not give an accurate estimate.

The position of the maximum of $P_{{\rm rad}}(s^{'})$ lies approximately
at
\begin{equation}
\sqrt{s^{'}} \approx m_{\pi} -
 \frac{\pi \beta}{8} \Gamma_{\pi},
\end{equation}
 which corresponds to a shift of about 0.2 eV from $m_{\pi}$,
and is therefore experimentally unobservable.

The differential decay rates $P_{0}$ and $P_{{\rm rad}}$ as functions of
the invariant mass of the electron-positron pair minus $m_{\pi}$, i.e.
$\sqrt{s^{'}} - m_{\pi}$, given by
 the numerical evaluation of the integral in Eq.(\ref{eq:integr}),
are shown in
Fig. \ref{fig:beauty}.
\begin{figure}[t]
\vspace{4.5in}
\caption{ The lowest order and radiatively corrected differential decay rates
 $ P_{0}(s^{'})$ (upper curve) and $ P_{{ \rm rad}}(s^{'})$
(lower curve) respectively,
normalized over $P_{0}(m^{2}_{\pi})$, as functions of
the electron pair invariant mass minus $m_{\pi}$, i.e.
$(s^{'})^{1/2} - m_{\pi}$. }
\label{fig:beauty}
\end{figure}
The effect of soft-photons radiated
in the final state is apparent, not only
 in the maximum height of the radiatively
corrected differential decay rate,
but also in the radiative "tail" that appears for lower
$e^{+}e^{-}$ invariant masses, as expected, since the photons are radiated
in the final state. Indicatively, numerical results show that the
radiatively corrected differential decay rate is roughly three times larger
than the non-corrected one at
$e^{+}e^{-}$ invariant masses 150 eV below $m_{\pi}$, and two times smaller
at 150 eV above $m_{\pi}$. In both cases, these differential decay rates
are roughly three orders of magnitude smaller than the ones at
 $s^{'} \approx m^{2}_{\pi}$.
 This situation is to be contrasted to the case of initial-state
radiative corrections, for processes like production of resonances
 via  $ e^{+}e^{-} $
collisions, where one sees a more pronounced radiative tail for invariant
masses
 larger than the resonance mass.

The experimental resolution of the invariant mass of the $e^{+}e^{-}$ pairs
being on the order of 0.5 MeV, the structure of the decay curve shown in
Fig.\ref{fig:beauty} is not experimentally observable. By numerically
integrating the function $P_{{\rm rad}}(s^{'})$ over $s^{'}$, it is
found that
the number of $e^{+}e^{-}$ pairs detected in an energy bin of width 0.5 MeV,
 centered around the pion mass, is approximately
$80 \%$ of the number it would be
without radiative corrections, i.e. we have
a $20 \%$ correction. This happens because
the radiative tail for $s^{'} < m^{2}_{\pi}$
  partly compensates  the reduced number of $e^{+}e^{-}$ pairs
at $s^{'} ~^{>}_{\sim} m^{2}_{\pi}$. A complete analysis of this process
should take into account this binning for energy regions further away
from $m_{\pi}$. However, results obtained in this way should not be trusted
for $\sqrt{s^{'}}$ too far away from $m_{\pi}$ (more than a few MeV), because
of several reasons. First,
because  background effects become non-negligible.
Moreover, in that
case, one should add to this calculation the contribution
of hard photons, which have an energy that is
not small compared to the electron
mass any more, and which become increasingly important for invariant masses
further away from $m_{\pi}$.  Last, the dependance of
the form factor,
 associated to the decay $\pi^{0} \rightarrow e^{+}e^{-}$, on $s^{'}$ cannot be
neglected any more.

It is furthermore expected that the effect of radiative
corrections on the total number of $e^{+}e^{-}$ pairs, having all possible
invariant masses, is dominated by perturbative quantum effects,
 and that these  are on the order of a few percent. In fact,
 a first-order in $\alpha$ prediction for this effect
(Eq.(18) in Ref.\cite{Bergrad}), which is a reasonable approximation,
 gives a $3 \%$ correction. It is therefore seen that the radiative
tail of the differential decay rate, when
considered throughout the whole range of allowable
$e^{+}e^{-}$ invariant masses, compensates, to within a few percent,
the reduction of the differential decay rate at $s^{'} \approx m^{2}_{\pi}$.
These radiative effects on the total number of  $e^{+}e^{-}$ pairs are not
going to be discussed further here, because they are completely
overwhelmed by interference terms between this process and its background,
 and by the background itself,
 which is discussed here below, and they are
 therefore experimentally unobservable.

To first order in the fine structure constant, the process considered here
involves a final state of an electron-positron pair and at most one
 photon. A large
background to this process is the same final state coming from the process
$\pi^{0} \rightarrow \gamma \gamma$, and the subsequent decay of one of the
final photons into an $e^{+}e^{-}$ pair \cite{Mikael}. This process is of
lower order in $\alpha$, and it has a much larger  decay
rate than the
one considered here. However, this background is easily distinguishable
from the signal of interest, because its differential decay rate does not have
a
pronounced peak at  $e^{+}e^{-}$ invariant masses $\sqrt{s^{'}}$
 near the pion mass. Moreover,
 the interference of this background with the process of interest in this paper
is negligible in the $\sqrt{s^{'}}$ region that is considered here, which is
 $\sqrt{s^{'}} \approx m_{\pi}$ (to within a few MeV) \cite{interf}.

To conclude, this paper has dealt with the shape of the
$\pi^{0} \rightarrow e^{+}e^{-}$ differential decay rate, when
 radiation of soft-photons from the two final electrons  is taken into account.
 In particular, the peak value of the radiatively corrected
differential decay rate $P_{{\rm rad}}$ is
smaller than the non-corrected one by about $50 \%$. Moreover,
the radiative tail, emerging at invariant masses $\sqrt{s^{'}}$
smaller than $m_{\pi}$, even
though it has a considerably smaller height than the peak differential decay
rate, it does not fall-off very rapidly with $\sqrt{s^{'}}$, and,
 when considered over the whole range of allowable electron invariant masses,
 it
almost compensates the reduction of the number of $e^{+}e^{-}$ pairs at
$\sqrt{s^{'}} \approx m_{\pi}$. This
form of the differential decay rate  should
be taken into consideration when comparison between experimental and
theoretical
results for this decay is made, especially when stringent
$e^{+}e^{-}$ invariant mass cuts
are placed.
A more involved analysis
should include the effect of higher order terms in the
quantities $\Delta^{1,2}_{{\rm rad}}$.
Finally, a similar analysis
can be done for the decay of other mesons, like the $\eta^{0}$, to a
lepton-antilepton pair.

\noindent {\bf Acknowledgements} \\
I thank Thomas Appelquist, Alan Chodos and Michael Zeller
for very helpful discussions.

\end{document}